\documentstyle[aps,twocolumn,epsfig]{revtex}
\begin{document}
\draft
\flushbottom
\twocolumn[
\hsize\textwidth\columnwidth\hsize\csname @twocolumnfalse\endcsname


\title{Absence of saturation for finite injected currents in
 axially symmetric cavity diode}
 
\vskip 0.2 in
\author{Debabrata Biswas, Raghwendra Kumar and R.~R.~Puri}
\address{
Theoretical Physics Division \\
Bhabha Atomic Research Centre \\
Mumbai 400 085, INDIA}
\date{\today}
\maketitle

\begin{abstract}

The Child-Langmuir law is investigated numerically using a fully
electromagnetic particle-in-cell code for a closed
axially symmetric diode. It is found that the average
current  transmitted to the anode ($J_{\rm TR}$) increases with the
injected current ($J_{\rm IN}$) even
after the formation of virtual cathode in both the non-relativistic
and relativistic cases. The increase is found to be a
power law, $J_{\rm TR}\sim J_{\rm IN}^{1-\beta}$. In other words, the time
averaged fraction $f$ of electrons reaching the anode varies
with the input current as, $f\sim J_{\rm IN}^{-\beta}$ where
$\beta < 1$. In contrast, for an infinite parallel plate diode,
$f\sim J_{\rm IN}^{-1}$. The possibility of asymptotic saturation
is also discussed.
\vskip .1 in
\end{abstract}

\date{today}
]
\narrowtext
\tightenlines

\newcommand{\be}{\begin{equation}}
\newcommand{\ee}{\end{equation}}
\newcommand{\bea}{\begin{eqnarray}}
\newcommand{\eea}{\end{eqnarray}}
\newcommand{\Tbar}{{\overline{T}}}
\newcommand{\En}{{\cal E}}
\newcommand{\Lop}{{\cal L}}
\newcommand{\DB}[1]{\marginpar{\footnotesize DB: #1}}
\newcommand{\q}{\vec{q}}
\newcommand{\kt}{\tilde{k}}
\newcommand{\Lopn}{\tilde{\Lop}}

\newpage
\section{Introduction}

	The Child-Langmuir law is the cornerstone of electron flow in
diodes. It gives the maximum current density, $J_{\rm CL}(1)$, that
can be transported from an infinite planar cold cathode 
(velocity of emission is zero) at zero potential
to an infinite planar anode parallel to the cathode at a distance $D$
and potential $V$. In the
non-relativistic case, this is given by \cite{child,langmuir}

\be
J_{\rm CL}(1) = {V^{3/2} \over D^2} {1 \over 9 \pi}
\left({2e\over m}\right)^{1/2}.
\label{eq:CL_1d}
\ee

\noindent
Here, $e$ is the magnitude of the charge and $m_0$ the rest mass of 
an electron.

	The Child-Langmuir law is a result of the space-charge effect.
As electrons are emitted from the cathode, they gain velocity
from the imposed field. At the same time, they experience
a repulsive force due to the presence of other electrons. However,
all the injected current is transmitted to the anode as long as
the injected current, $J_{\rm IN}$, is less than a particular value,
called the critical current $J_{\rm CR}$. When $J_{\rm IN}$ exceeds
$J_{\rm CR}$, the net force is such that some of the electrons,
instead of moving towards the anode, start moving backwards as if
reflected from a ``virtual cathode''. As a result, only a part of the
current, $J_{\rm TR}$, is transmitted. The position of the
virtual cathode depends on $J_{\rm IN}$ and on the 
initial velocity, $v_0$, of the emitted electrons.
In one dimension, $J_{\rm CR}=J_{\rm CL}(1)$ in case of cold
emission i.e. in case $v_0=0$. Also,  the virtual cathode is formed on
the cathode as soon as the injected current exceeds $J_{\rm CR}$ and
remains there even when $J_{\rm IN}$ is increased. The transmitted
current $J_{\rm TR}$ in this case is the Child-Langmuir current
$J_{\rm CL}(1)$ for any value of the injected current above
$J_{\rm CR}$. The transition in the transported current is therefore
sharp: $J_{\rm TR} = J_{\rm IN}$ if $J_{\rm IN}\le J_{\rm CR}$,
$J_{\rm TR} = J_{\rm CR}$ if $J_{\rm IN}\ge J_{\rm CR}$.

	The flow of current below the critical value is steady i.e.
time-independent, but is oscillatory above it. The value of the
critical current $J_{\rm CR}$ is thus that value above which the
system does not have a steady state. This criterion may be used to
evaluate exact analytical expression for $J_{\rm CR}$ in
one-dimensional planar geometry in case the kinetic energy of the
electrons is non-relativistic. The analytical expression for
$J_{\rm CR}$ may also be derived in the ultra-relativistic limit.

	Since, as mentioned above, the system does not approach any steady
state if $J_{\rm IN}>J_{\rm CR}$, its behaviour for the input currents
exceeding the critical current is described by time-dependent
equations. However, it is not possible to solve the time-dependent
equations analytically exactly even in one-dimensional
planar geometry \cite{bb}. Solving time-dependent equations is avoided in the
so called classical theory which, for $J_{\rm IN}>J_{\rm CR}$, assumes
a steady state which takes account of reflections from the virtual
cathode. The one-dimensional planar model is analytically exactly
solvable within the framework of the classical theory. In particular,
it leads to the Child-Langmuir law. Comparison with numerical work
shows that the predictions of the classical theory are close to the
time-averaged values of physical quantities.

	Over the past decade, several studies have been carried
out to extend the Child-Langmuir law to two and higher dimensions 
\cite{CL2D,CL2D1}. 
The motivation for this is twofold. In the first place, this helps to
benchmark particle-in-cell (PIC) simulation codes. More importantly,
the existence of a space charge limited saturation current implies
that one need not bother about the material dependent cathode
characteristics (the Richardson-Dushman law in case of thermionic
emission or the Fowler-Nordheim law \cite{FN} for field emission) 
so long as the injected current is more than what gives rise to 
the limiting value of
the transmitted current. This condition considerably simplifies the
simulation of devices that use a diode.

	Most studies in two-dimensions have centred around 
cold emission from
a finite emission area in an otherwise infinite parallel plate
geometry \cite{CL2D,CL2D1,CL2D2}. 
Note that the Child-Langmuir current is defined in these
studies as the current at which oscillatory behaviour sets in
(the critical current, $J_{\rm CR}$ ). Both numerical and analytical
results indicate that this critical current in two-dimensions
($J_{CL}(2)$) is higher than the Child-Langmuir current in
one-dimension ($J_{CL}(1)$). Further, there exists a relation between
the two \cite{CL2D1}:

\be
{J_{\rm CL}(2) \over J_{\rm CL}(1)} = 1 + a_0 {D \over W}.
\label{CL2}
\ee

\noindent
Here $W$ is width of the emission strip, $D$ is the anode-cathode
separation and $a_0 \simeq 0.31$. The question of
limiting current (the maximum
current transported across the gap) has not however been
investigated by these authors.

Recent experiments using photoemission from a cathode have yielded
some interesting results \cite{valfells}. The experiments show that it is
possible to transport a much higher current than that predicted
by the one-dimensional law for a short emission pulse or limited
emitter area and (more importantly) the current may be increased
even after formation of the virtual cathode. This leads to a
distinction between the critical current (the minimum current
required for virtual cathode formation) and the limiting current.
In the range of laser intensity studied in these experiments, no
limiting current (maximum current that can be transported for a given
potential difference and gap separation) was established although
the authors suggest the existence of an asymptote.

The question of limiting current in two and higher dimensions thus
assumes a great significance since its non-existence in realistic
situations can be a major stumbling block in simulations. With this in
mind, in this communication we investigate numerically the
Child-Langmuir law in an axially symmetric cavity. We demonstrate
that, for axially symmetric closed geometries with a finite emitting
area, (i) the current increases even after formation of the virtual
cathode, and (ii) the time-averaged fraction of electrons reaching the
anode varies as $f \sim J_{\rm IN}^{-\beta}$ with $\beta < 1$ thereby
establishing that the transmitted current, $J_{\rm TR}=fJ_{\rm IN}$,
increases as a power law beyond the critical current.

The plan of the paper is as follows: We first review the
one-dimensional non-relativistic case and, within the framework of the
classical theory, show analytically that above the critical current,
the fraction of charges reaching the anode varies inversely with the
injected current ($f \sim J_{\rm IN}^{-1}$).
A similar study for the relativistic case is presented next
followed by our numerical results for infinite planar diode and
axially symmetric cavity diode. Finally our results are summarized 
in the concluding section. As a matter of convenience, we shall use
the notations $J$, $v$ and $\rho$ to refer to the {\em magnitude} of
the current density, velocity and charge density respectively. 

\section{Critical and Limiting currents: one-dimensional
non-relativistic case}
\label{sec:limiting}

\subsection{Determination of Critical Current}

Consider two infinite plates placed at $z=0$ and $z=D$ at fixed
potentials given by
\be
\phi(z=0)=0, \qquad \phi(z=D)=V.
\label{bc}
\ee
Assume that the system evolves to a steady state. In that state, the
current at every plane between the plates approaches a value which is
independent of time. It can be shown that the potential $\phi(z)$ at
the plane at the position $z$ in that case is the solution of the
Poisson equation
\be
{{\rm d}^2 \phi \over {\rm d}z^2} = -4\pi(-\rho(z)) = 4\pi \rho(z)
\label{eq3}
\ee

\noindent
in which the charge density $\rho(z)$ is determined by the condition
that the current density is independent of space so that
$\rho(z) v=J_{\rm IN}$ whereas the velocity $v$ is determined by the
energy conservation law $\En_0 = {1\over 2} m_0 v^2 - e\phi(z)$ where
$\En_0 = {1\over 2} m_0 v_0^2 - e\phi(0)$ is the initial energy of an
electron at the surface of the cathode. Under the condition of cold
emission, $v_0=0$, and due to the boundary condition (\ref{bc}) at
$z=0$, it follows that $\En_0=0$. Hence, the law of energy
conservation in the steady state reads
\be
\En = {1\over 2} m_0 v^2 - e\phi(z)=0.
\label{ec}
\ee
Equation (\ref{eq3}) then assumes the form
\be
{{\rm d}^2 \phi \over {\rm d}z^2} = {C\over \sqrt{\phi}},
\label{eq4}
\ee
\noindent
where
\be
C=4\pi J_{\rm IN}(m/2e)^{1/2}.
\label{ceqt}
\ee
Equation (\ref{eq4}) is to be solved under the boundary conditions
given in (\ref{bc}).

The steady state may alternatively be found by solving the equation of
motion for electrons \cite{akimov}. 
That equation, in the Llewellyn form \cite{bb,llewellyn}, reads

\be
{{\rm d}^3 z(t,t_0) \over {\rm d}t^3} = - {e\over m_{0}} J_{T}(t)
\label{eq:Llewellyn}
\ee

\noindent
where $t_0$ is the time at which the electron is injected
(initial time), $m_{0}$ the rest mass of an electron,
$J_{T} = \partial E_0(t)/\partial t + 4\pi J$ is the total current
and $E_0$ is the electric field on the surface of the cathode.

Eq.~(\ref{eq:Llewellyn}) can be integrated to get

\bea
z(t,t_0) & = &  v_0(t-t_0) - {e\over 2m_0} E_0(t_0)(t-t_0)^2 \nonumber \\
& - & {e\over 2m_0} \int_{t_0}^t d\tau (t-\tau)^2 J_T(\tau).
\eea


\noindent
In the steady state, $E_0(t) = E_0(t_0)$ and
$\partial E_0/\partial t =0$. Also, using the fact that below the
critical current, no electron is reflected back, we have
$J =J_{\rm IN}$. Thus,

\bea
z(t,t_0) &  = & v(t-t_0) - {e\over 2m_0} E_0(t_0)(t-t_0)^2 \nonumber \\
& + & {4\pi eJ_{\rm IN}\over 6m_0} (t-t_0)^3,
\label{eq:z0}
\eea


\noindent
and

\be
v(t,t_0)  =  v_0 - {e\over m_0} E_0(t_0)(t-t_0) 
 +  {4\pi eJ_{\rm IN} \over 2m_0} (t-t_0)^2.
\label{eq:vz0}
\ee

\noindent
For the case of cold emission ($v_0=0$), (\ref{ec}) holds giving
$v = (2e\phi/m_0)^{1/2}$ so that the final velocity at the anode
is $v_D = (2eV/m_0)^{1/2}$, where $\phi(D) = V$.
Let $T$ denote the time that an electron takes to transit the
cathode-anode distance ($D$). Then, with  $z=D$, $v=v_D$, $t-t_0=T$,
Eqs. (\ref{eq:z0}) and (\ref{eq:vz0}) assume the form

\bea
L & = & -{e\over 2m_0} E_0 T^2 + {4\pi eJ_{\rm IN}
\over 6m_0} T^3 \label{eq:z1} \\
v_D & = & {e\over m_0} E_0 T  + {4\pi eJ_{\rm IN} \over 2m_0} T^2
\label{eq:vz1}
\eea

\noindent
These equations determine the two unknowns, $E_0$ and $T$.
On eliminating $E_0$,
the equation for T in terms of the scaled transit time $\Tbar $ is

\be
\Tbar^3 - \alpha  \Tbar + \alpha=0
\label{cubic}
\ee

\noindent
where $\alpha = 27 V^{3/2}/4\beta$,
$\Tbar = T/T_0$, $T_0 = D/(2m_0/eV)^{1/2}$ and
$\beta = 9 \pi J_{\rm IN} D^2 (m_0/2e)^{1/2}$. Note that $T_0$ is the
transit time in the absence of space charge. The condition that all
the roots of (\ref{cubic}) be real may be shown to be $\alpha > 27/4$.
For $\alpha < 27/4$, two of the roots are complex while one root is
real. The real root is however negative and hence inadmissible as a
transit time. Thus $\alpha = 27/4$ marks the critical current. On
substituting the values of $\alpha$ and $\beta$, the expression for
the critical current is found to be given by

\be
J_{\rm CR} = {V^{3/2} \over D^2} {1\over 9\pi}
\left({2e \over m_0}\right)^{1/2}.
\label{cr}
\ee
\noindent
Also, note that at $J_{\rm IN}=J_{\rm CR}$, $\Tbar = 3/2$ and hence
$E_0 = 0$. Thus, at the critical current, the electric field at the
cathode vanishes.

\subsection{Determination of Limiting Current : Classical Theory}

The above analysis holds good for $J_{\rm IN} \leq J_{\rm CR}$ where the
injected and transmitted currents are equal. Beyond the critical
current, electrons are reflected and the analysis breaks down. In
order to find out what happens beyond the critical current, it is
useful to consider yet another steady state model which assumes that
a fraction $f$ of the injected electrons is transmitted while the
fraction $1-f$ is reflected at some point $z=z_m$ so that the 
transmitted current is $J_{\rm TR} = f J_{\rm IN}$ \cite{langmuir,drift}. 
Though we have
assumed so far that the total energy $\En_0$ of an electron entering the
diode is zero, it is instructive to carry the classical theory for
injected currents greater than the critical currents for the case of
non-zero $\En_0$. To that end, on invoking the discussion following
Eq.(\ref{eq3}), the Poisson equation in the two regions,
$0\le z\le z_m$ and $z_m\le z\le D$, then assumes the form

\bea
{\rm d^2\bar\phi\over{\rm d}\bar z^2}&=
&{4\alpha (2-f)\over 9\sqrt{\bar\phi}},~~~0\le z\le z_m,
\label{eq:poisson_L} \\
{{\rm d}^2\bar\phi\over{\rm d}\bar z^2}&=&
{4\alpha f\over 9\sqrt{\bar\phi}},~~~z_m\le z\le D,
\label{eq:poisson_R}
\eea
where $\bar\phi=(\phi+\En_0/e)/V$, $\bar z=z/D$, and
\be
\alpha = {9 \pi J_{\rm IN} D^2 \over V^{3/2} (2e/m_0)^{1/2}}.
\label{alpha}
\ee

\noindent
Both these equations are of the form
${\rm d}^2\bar\phi/{\bar d}\bar z^2 = A/\bar\phi^{1/2}$ and can be
cast in the form
\be
{{\rm d}\over {\rm d}\bar z}\left({{\rm d}\bar\phi\over
{\rm d}\bar z}\right)^2 =4A {{\rm d}\sqrt{\bar\phi}\over{\rm d}\bar z}.
\ee

\noindent
The solution of this equation reads

\be
\left({{\rm d}\bar\phi\over{\rm d}\bar z}\right)^2 = 4A \sqrt{\bar\phi}
 + B
\label{eq:poisson1}
\ee

\noindent
Since the electrons come to a stop at $\bar z=\bar z_m$, the law
of conservation of energy ($\En_0 = m_0 v^2/2 - e\phi$) 
implies that $\bar\phi (\bar z_m) = 0$.
Also, the field on the electrons at $z=z_m$ is assumed to be zero.
Hence
\be
{{\rm d}\bar \phi\over{\rm d}\bar z}\Big|_{\bar z=\bar z_m} = 0.
\ee
On using the abovementioned values of $\bar\phi$ and
${\rm d}\bar\phi/{\rm d}\bar z$ at $\bar z_m$, it follows from
Eq.~(\ref{eq:poisson1}) that $B=0$.  Thus

\be
{{\rm d}\bar\phi\over{\rm d}\bar z} = \pm 2\sqrt{A}\bar\phi^{1/4},
\label{eq19}
\ee

\noindent
where minus sign holds in the region $0\le\bar z\le \bar z_m$
whereas the plus sign is for the region $\bar z_m\le\bar z\le 1$.
That is because $\bar\phi(z)$ decreases from $\bar z=0$ to
$\bar z=\bar z_m$ so that the electrons are decelerated and
increases beyond $z_m$ so that a fraction of electrons is
transmitted after they come to rest at $\bar z=\bar z_m$. The
solution of (\ref{eq19}) yields

\be
\bar \phi^{3/2} = {9\over 4} A (\bar z - \bar z_m)^2.
\ee
On using the appropriate values of $A$ in the two regions, the
solution can be expressed as

\bea
\bar\phi^{3/2} & = & \alpha (2-f)(\bar z-\bar z_m)^2,
~~~0 \leq \bar z \leq \bar z_m, \nonumber \\
\bar\phi^{3/2} & = & \alpha f (\bar z - \bar z_m)^2,
~~~\bar z_m \leq \bar z \leq 1.
\label{eq:poi_sol_R}
\eea
Now, on applying the boundary conditions (\ref{bc}), the equations
in (\ref{eq:poi_sol_R}) yield

\bea
\left({\En_0\over eV}\right)^{3/2}&=&\alpha (2-f)\bar z^2_m,\nonumber\\
\left({V + \En_0/e\over V} \right)^{3/2}&=&\alpha f(1-\bar z_m)^2.
\label{nen1}
\eea
These equations show that if,
$\En_0=0$ then
\be
z_m=0~~{\rm and}~~\alpha f = 1.
\label{eqn2n}
\ee

The position of the virtual cathode is thus on the cathode when the
initial energy of an electron is zero but is away from it otherwise.
On substituting for $\alpha$ from (\ref{alpha}) in (\ref{eqn2n}),
the transmitted current is found to be given by
\be
J_{\rm TR} = fJ_{\rm IN} = {V^{3/2} \over D^2} {1\over 9\pi}
\left({2e \over m_0}\right)^{1/2}.
\label{lc(1)}
\ee
\noindent
Note that this value of the transmitted current is independent of the
injected current and that this expression holds as long as the
injected current exceeds the critical current.
Eq.~(\ref{lc(1)}) is thus the limiting current that can flow through
the diode. Note also that the limiting current (\ref{lc(1)}) 
is identical to the critical current $J_{\rm CR}$ given in
(\ref{cr}).

It should be emphasized that the steady state model for
$J_{\rm IN} > J_{\rm CR}$ is
phenomenological. As is revealed by the numerical solution of the
time-dependent equations for injected currents above $J_{\rm CR}$,
the behaviour of the system is oscillatory i.e. it does not approach
a steady state. However, the steady state values derived above are
found to be in close agreement with the time-averaged values of the
numerical results.

\section{Critical and Limiting Currents: one-dimensional relativistic
case}
\label{sec:rel}

Finding an explicit expression for the critical current in case of
relativistic electronic energies is generally a formidable task.
However, an analytical expression for the critical current may be
derived in the ultrarelativistic limit in case the initial velocity
of the electrons is zero if, following the results of the
non-relativistic treatment, we identify the critical current for zero
initial velocity as the one for
which the electric field at the cathode is zero. We will see that,
as in the non-relativistic limit, this current is the same as the
limiting current predicted by the classical theory for relativistic
energies.

\subsection{Critical Current}

The expression for the critical current in the relativistic case
may be derived more conveniently by starting from the Poisson
equation (\ref{eq3}). The charge density  $\rho$ in (\ref{eq3}) is
determined by $\rho v=J_{\rm IN}$ and $v$ by the relativistic
energy conservation law
\be
m_0c^2 (\gamma(z)-1) - e\phi(z) = 0,
\label{eq27}
\ee
\noindent
where $\gamma(z) = (1 - v^2(z)/c^2)^{-1}$. Eq.~(\ref{eq3}) may
then be rewritten as
\be
{{\rm d}^2 \gamma(z)\over {\rm d}z^2}
= {K \gamma  \over \sqrt{\gamma^2(z) - 1} }
\label{eq:poi_rel2}
\ee

\noindent
where $K = 4\pi J_{\rm IN} (e/m_0 c^3)$. The boundary conditions (\ref{bc})
in this case assume the form
\be
\gamma(z=0)=1,~~~~~~~~~\gamma(z=D)=1+{eV\over m_0c^2}\equiv \gamma_D.
\label{bcr}
\ee

Following the non-relativistic
case, we assume that when the initial velocity of the electrons is zero,
the critical value of transmitted current is attained when the field
at cathode vanishes. Hence, assuming ${\rm d}\phi(z)/{\rm d}z =0$ at
$z=0$ and invoking also the boundary condition (\ref{bcr}) at $z=0$,
the integration of (\ref{eq:poi_rel2}) gives
\be
\left({{\rm d}\gamma(z)\over{\rm d}z}\right)^2 =
2K(\gamma^2 - 1)^{1/2}.
\label{eq29}
\ee
On integrating this equation with the boundary condition (\ref{bcr})
at $z=0$ it may be shown that the current density (which, as
explained above, is to be identified as the critical current density)
is given by
\be
J_{\rm CR}={m_0 c^3 \over 8\pi ez^2}I^2(\gamma),
\label{eq32}
\ee
where
\be
I(\gamma) =
\int_1^\gamma \gamma^{-1/2}(1 - \gamma^{-2})^{-1/4} {\rm d}\gamma.
\label{eq33}
\ee
Eq.(\ref{eq32}) determines $\gamma$ implicitly as a function of
$z$ and $J_{\rm CR}$ in case the field at the cathode is zero.
However, $J_{\rm CR}$ is still an unknown. It may be determined by
applying the yet unused boundary condition at $z=D$ leading to the
relation 
\be
J_{\rm CR}={m_0 c^3 \over 8\pi eD^2}I^2(\gamma_D),
\label{eq32n}
\ee
where $\gamma_D$ is given by (\ref{bcr}) in terms of the applied
voltage $V$. Hence, (\ref{eq32n}) determines the critical current in
terms of known quantities.

The function $I(\gamma)$ in (\ref{eq33}) may be evaluated as follows:
\bea
I(\gamma)& = & \int_1^\gamma \gamma^{-1/2}(1 - \gamma^{-2})^{-1/4}
{\rm d}\gamma\nonumber \\
& = & \int_1^\gamma{\rm d}\gamma \sum_{m=0}^{\infty}
{\Gamma\left(m+{1\over 4}\right)
\gamma^{-2m-1/2} \over \Gamma(1/4) m!} \nonumber\\
& = & {1\over \Gamma(1/4)}
\sum_{m=0}^{\infty}\int_1^\gamma{\rm d}\gamma
{\Gamma\left(m+{1\over4}\right) \over m!} \gamma^{-2m-{1\over 2}} 
\nonumber \\
& = & -{1\over 2}{\Gamma(1/4)\over\Gamma(3/4)}
\big[\gamma^{1/2} \tilde{F}(\gamma^{-2}) -
\tilde{F}(1) \big]
\label{eq:integral}
\eea

\noindent
where $\tilde{F}(x) \equiv F({1\over 4},-{1\over 4},{3\over 4};x)$
stands for the Hypergeometric function. Now, let $\gamma=\gamma_D$
and let the voltage be such that $eV>>m_0c^2$ so that
$\gamma_D>>1$. Hence, $\tilde{F}(\gamma_D^{-2})\approx 1$. Using

\be
F(a,b,c;1)={\Gamma(c-a-b)\Gamma(c)\over \Gamma(c-a)\Gamma(c-b)},
\ee
and $\Gamma(3/4) = 1.2254167024$ \cite{AS}, 
the expression for
the critical current in the ultrarelativistic limit turns out to be
given by
\be
J_{\rm CR} = {m_0 c^3 \over 2\pi e D^2}\left[\left(1 +
{eV \over m_0 c^2}\right)^{1/2} - 0.8471\right]^2.
\label{eq34}
\ee
\noindent
This is in agreement with the result of Jory and Trivelpiece \cite{jory}
derived under identical boundary conditions though these authors do not
identify (\ref{eq34}) as the critical current.

\subsection{Limiting Current}

As in the non-relativistic case, we assume that when
$J_{\rm IN} > J_{\rm CR}$, the steady state is characterized by a
fraction $f$ of particles transmitted beyond the position $z_m$ at
which the velocity becomes zero so that the current in the region
between the cathode and the position $z_m$ consists of two parts, the
injected current density $J_{\rm IN}$ moving away from the cathode to
the virtual cathode at $z_m$ and the reflected part $(1-f)J_{\rm IN}$
moving to the cathode from the virtual cathode. The Poisson
equation in the two regions can then be written as

\bea
{{\rm d}^2 \gamma\over {\rm d}z^2} & = & {(2-f)K\gamma\over
\sqrt{\gamma^2 - 1}},~~~0 \leq z \leq z_m,
\label{eq:poisson_rel_L} \\
{{\rm d}^2 \gamma \over {\rm d}z^2} & = & {f K\gamma\over
\sqrt{\gamma^2 - 1}},~~~z_m \leq z \leq D,
\label{eq:poisson_rel_R}
\eea

\noindent
where we have assumed that the energy of the injected electrons
is zero. Energy conservation thus leads to the condition 
$(\gamma - 1)m_0c^2 - e\phi = 0$.
The two equations above are of the same form, whose solution is

\be
\left({{\rm d}\gamma(z) \over {\rm d}z}\right)^2 =
2\tilde{K}(\gamma^2 - 1)^{1/2} + C
\ee

\noindent
where $\tilde{K} = (2-f)K$ for $0 \leq z \leq z_m$ and $\tilde{K}=f K$
for $z_m \leq z \leq D$. Note that, at $z=z_m$, $\gamma = 1$,
${\rm d}\gamma(z)/{\rm d}z=0$. Thus $C=0$.
Keeping in mind that the potential decreases in ($0,z_m$) and
increases in ($z_m,D$), the solutions in the two regions read:

\bea
\int_1^\gamma {{\rm d}\gamma \over (\gamma^2 - 1)^{1/4} } & = &
(2(2-f)K)^{1/2}(z_m-z),~~~z\le z_m, \nonumber \\
\int_1^\gamma {{\rm d}\gamma \over (\gamma^2 - 1)^{1/4} } & = &
(2fK)^{1/2}(z-z_m), ~~~z\ge z_m.
\label{eq38}
\eea

\noindent
Like in the non-relativistic case discussed in the last section, the
analysis may be carried for non-zero values of initial energy $\En_0$.
Restricting, however, to the case of $\En_0=0$ with zero initial
velocity (so that $\gamma(z=0) = 1$) and zero potential at the
cathode, the boundary conditions give $z_m = 0$ and
\be
fJ_{\rm IN}={m_0 c^3 \over 8 z^2\pi e}I^2(\gamma),
\label{eq39}
\ee

\noindent
where $I(\gamma)$ is as in Eq.~(\ref{eq33}). Eq.(\ref{eq39}) with
$fJ_{\rm IN} \rightarrow J_{\rm CR}$ is the same as (\ref{eq32}). On applying
the boundary condition $\gamma=\gamma_D$ at $z=D$ we get the current
at the anode. Hence, the transmitted current $fJ_{\rm IN}$ is the same
irrespective of the value of the injected current $J_{\rm IN}$ as long 
as $J_{\rm IN}$ is above the critical value at which virtual cathode 
is formed. The
critical value of the current is, therefore, the same as the limiting
current. Note from (\ref{eq39}) that the fraction of electrons
transmitted varies as $f \sim J_{\rm IN}^{-1}$.

\section{Numerical Results}

	The 1-D electrostatic analysis reviewed above indicates that, for
cold emission, the critical and limiting currents are
identical in both non-relativistic and relativistic cases. The chief
motivation of this study is to
investigate whether this holds in higher dimension. To this end,
we shall use
the fully electromagnetic particle-in-cell code, SPIFFE \cite{spiffe},
and limit ourselves to axially symmetric diodes.
For completeness and comparison, we shall first present our
numerical results for 1-dimension.

\subsection{One-Dimension: Sheet Model}

	In practical terms, the 1-dimensional analysis applies when
the cathode and anode are parallel plates of dimensions much larger
than the separation between them with uniform emission from the
surface of the cathode. A description, convenient for numerical
calculations, is to discretize continuous electronic fluid as sheets
of uniform charge density parallel to the surface of the two
electrodes \cite{bb}.
The sheets move in the direction perpendicular to their
surface. The position $z_k$ of the ${\rm k}^{{\rm th}}$ sheet 
at time $t$
is governed by the equation \cite{bb_sheet}

\be
{{\rm d} \beta_k \over {\rm d}t} =
{2\over 9}{\alpha_{\rm rel} \over \gamma_k^3 N}
\big[\sum_{i=1}^{M} \overline{z}_i - k - {1\over 2} \big].
\ee

\noindent
This equation takes into account the electrostatic repulsion between
the sheets. Here $\overline{z}_k = z_k/D$ is the scaled position of
the $k^{{\rm th}}$ sheet, $\beta_k = v_k/c$, $\alpha_{{\rm rel}}
= 4\pi J_{\rm IN} D^2 (9e/2m_0c^3) (c/v_0)^3$, $v_0$ is the initial velocity,
$N$ is the number of sheets launched per unit transit time in the
absence of space charge and $M$ is the total number of
sheets present in the diode at any instant of time.

For the numerical calculations presented here, the integration
time step $\Delta t$ is $D/(200 v_0)$, $D=0.8$ cm while $N=2000$ for
non-relativistic calculations ($V=250$ kV) and $N=10000$ for relativistic
calculations ($V=2$ MV). These parameters were chosen to satisfy
convergence requirements.

As stated before, the system approaches a steady state if the injected
current is below its critical value whereas the solution is
oscillatory for currents above that value. Since the classical
theory is based on the assumption of a steady state even above the
critical current, we compare it with the time-averaged behaviour of
the physical quantities if the injected current is above the
critical current \cite{transient}.

Fig.~1 is a plot of the transmitted current averaged over 10ns
as a function of the injected current for $V = 250$ kV. It is clear
that $J_{\rm TR}$ attains saturation at $J_{\rm IN}=4.3~{\rm MA/m^2}$. 
In order to quantify the saturation,
we have studied the (time averaged) fraction $f$ of electrons reaching the
anode as a function of the injected current. According to the
analysis presented in section \ref{sec:limiting},
$f \sim J_{\rm IN}^{-1}$.  Fig.~2 confirms
that $f \sim J_{\rm IN}^{-\beta}$ with $\beta = 0.999$

\begin{center}
\begin{figure}[tbp]
\hspace*{0.1cm}\epsfig{figure=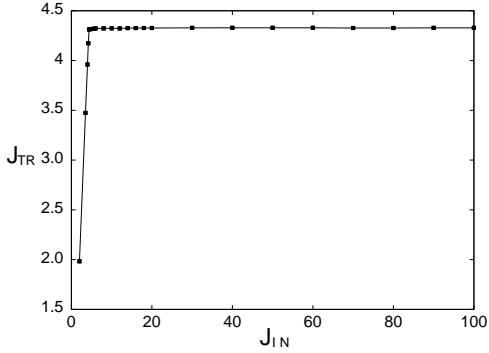,width=4.5cm,angle=270}
\vskip 0.25 in
\caption[ty] {Time averaged transmitted current density plotted
as a function of 
the injected current density for the 1-dimensional case. 
Both are measured in ${\rm MA/m^2}$.
Note the sudden transition at $4.3$ where saturation sets in. The theoretical
value using Eq.~(\ref{eq:CL_1d}) is 4.5.}
\label{fig:1}
\end{figure}
\end{center}

\begin{center}
\begin{figure}[tbp]
\hspace*{0.1cm}\epsfig{figure=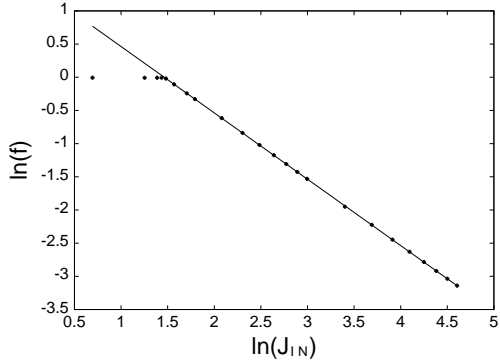,width=4.5cm,angle=270}
\vskip 0.25 in
\caption[ty]{A log-log plot of the average fraction of transmitted
electrons versus the injected current for the 1-dimensional case. 
The linear fit  confirms that
$f \sim J_{\rm IN}^{-\beta}$ with $\beta = 0.999$.}
\label{fig:2}
\end{figure}
\end{center}

\begin{center}
\begin{figure}[tbp]
\hspace*{0.1cm}\epsfig{figure=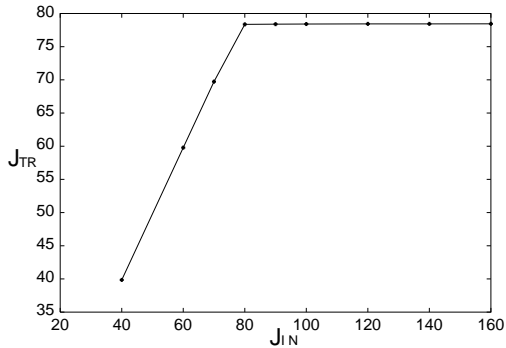,width=4.5cm,angle=270}
\vskip 0.25 in
\caption[ty]{Time averaged transmitted current density plotted as a 
function of the injected current density for the 1-dimensional 
relativistic case. Both are measured in ${\rm MA/m^2}$.
The critical current density is found to be $78.3~{\rm MA/m^2}$ after which
saturation sets in. The theoretical value using (\ref{eq34}) is 
$80~{\rm MA/m^2}$. }
\label{fig:3}
\end{figure}
\end{center}

A similar study of the 1-dimensional relativistic case (2 MV)
leads us to an identical conclusion - the critical and limiting
currents are indeed identical. Fig.~3 provides confirmation of this.
Note that the ultra-relativistic analysis, 
presented in section \ref{sec:rel}
predicts $J_{\rm CR} \simeq 80~{\rm MA/m^2}$ 
while the observed value is $78.3~{\rm MA/m^2}$. The
saturation has also been studied using the averaged fraction of
transmitted electrons as a function of the injected current. It is
found that $f \sim J_{\rm IN}^{-0.998}$.

	Thus, in the  1-dimensional case, the critical and limiting
currents are indeed identical and, above the critical current, the
fraction of electrons transmitted varies as $f\sim J_{\rm IN}^{-1}$.

\subsection{Axially Symmetric Diode: PIC}

	For the axially symmetric diode, the numerical calculations
were performed using the fully electromagnetic particle-in-cell
code SPIFFE \cite{spiffe}. The basic algorithm is described below 
\cite{borland}. 

\begin{enumerate}

\item
 Create the computational mesh and assign metal or vacuum points
according to the geometry specified in the input file.

\item
 Evaluate imposed fields due to potentials applied on metal surfaces

\item
 Inject macroparticles with specified charge and velocity using the
``over-injection'' method.

\item
 Distribute the charge and velocity to the mesh points using a 
standard weighting scheme.

\item
 Solve Maxwell's curl equations with the specified
boundary conditions (Dirichlet/Neumann) using the Finite Difference
Time Domain (FDTD) method.

\item
   Find forces at the position of the particles by interpolating 
the field values at the adjacent grid points.

\item
 Using these forces, find new position and momentum of the particles.

\item
 Remove particles that reach the end of the simulation region or hit 
a metal surface.

\item Repeat 1 to 8.

\end{enumerate}

 After a specified number of time steps, it checks if
Poisson's equation is satisfied within a specified ``error charge''. 
If not, it corrects the electric field after solving Poisson's equation
using the error charge.

\begin{center}
\begin{figure}[tbp]
\vskip -0.25 in
\hspace*{0.1cm}\epsfig{figure=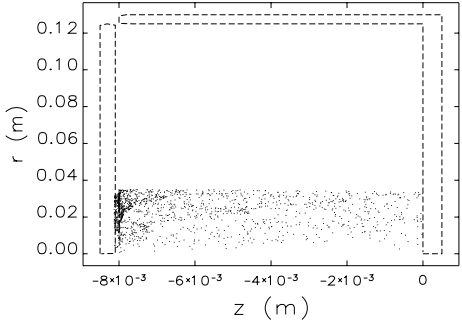,width=5cm,angle=270}

\vskip -0.05 in
\caption[ty] {The geometry of the diode and a typical plot of the
particles for $J_{\rm IN} > J_{\rm CR}$. Here $J_{\rm IN} = 6 {\rm MA/m}^2$
and  $V=250$ kV. 
}

\label{fig:geometry}
\end{figure}
\end{center}

The geometry of the diode consists of a hollow cylinder
of radius 12.5 cm 
with the cathode and anode plates of radius 12.5 cm placed on either end
with a separation $D = 0.8$ cm between them.
The cylinder and anode plate are connected and maintained 
at a potential, $V$. The cathode plate (on the left in 
fig.~\ref{fig:geometry})
is grounded and the emitting area is restricted
to a radius R=3.5 cm. In each case, 
the integration time step was chosen to be $0.0002$ ns. Convergence
was checked against the mesh spacings in $r$ and $z$ 
as well as the charge per macro-particle.

\begin{center}
\begin{figure}[tbp]
\vskip -0.25 in
\hspace*{0.1cm}\epsfig{figure=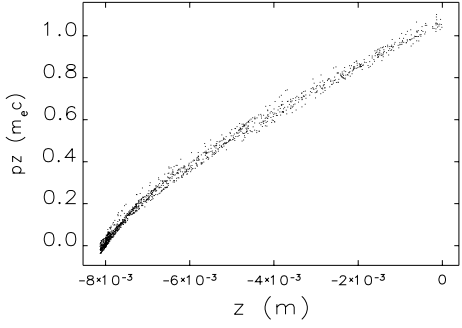,width=5cm,angle=270}

\vskip -0.05 in
\caption[ty] {The phase space plot corresponding to fig.~4. Note that $p_z$ is
measured in units of $m_0 c$. In the absence of space charge,
$p_z = 1.1$ at the anode for $V = 250$~kV.
}

\label{fig:phase_space}
\end{figure}
\end{center}

\begin{center}
\begin{figure}[tbp]
\vskip -0.25 in
\hspace*{0.1cm}\epsfig{figure=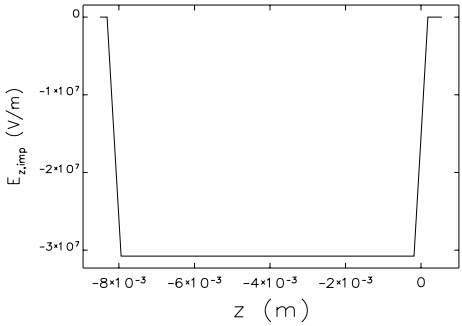,width=5cm,angle=270}

\vskip -0.05 in
\caption[ty] {The imposed electric field, $E_{z,imp}$ 
due to the applied potential
}
\label{fig:EzImposed}
\end{figure}
\end{center}

The simulation geometry along with a typical plot of the particles
in configuration space for $J_{\rm IN} > J_{\rm CR}$ 
is shown in fig.~\ref{fig:geometry}. 
The phase space plot corresponding to fig.~\ref{fig:geometry}
is shown in fig.~\ref{fig:phase_space}.

\begin{center}
\begin{figure}[tbp]
\vskip -0.25 in
\hspace*{0.01cm}\epsfig{figure=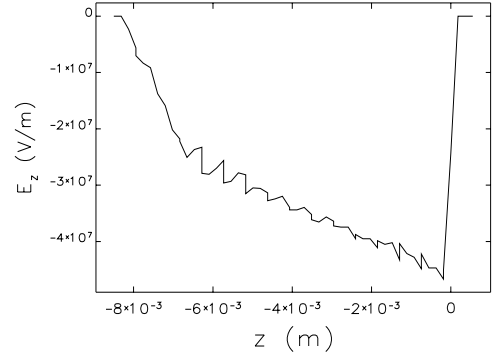,width=5cm,angle=270}

\vskip -0.05 in
\caption[ty] {The electric field after 2 ns. Note that the field
at the cathode is nearly zero.
}
\label{fig:Ez}
\end{figure}
\end{center}

\begin{center}
\begin{figure}[tbp]
\hspace*{0.01cm}\epsfig{figure=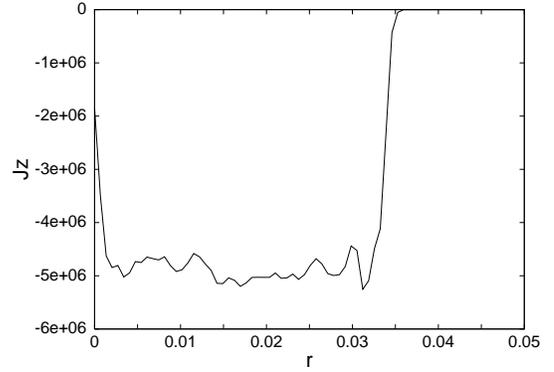,width=5cm,angle=270}

\vskip 0.05 in
\caption[ty] {The time averaged current density 
close to the anode plotted against the radius. 
Note that the emitter radius was 0.035 m.
}
\label{fig:Jz_radial}
\end{figure}
\end{center}

\begin{center}
\begin{figure}[tbp]
\hspace*{0.01cm}\epsfig{figure=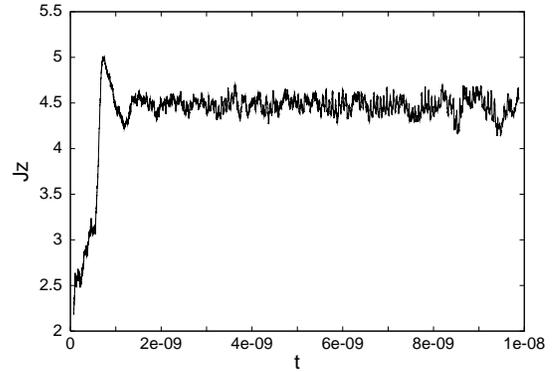,width=5cm,angle=270}

\vskip 0.05 in
\caption[ty] {The locally time averaged current density 
as a function of time. Note that it is steady after the 
initial transients. 
}
\label{fig:Jz_t}
\end{figure}
\end{center}

Figs.~\ref{fig:EzImposed} and \ref{fig:Ez} are plots of the 
electric field $E_z$ at the beginning of the simulation 
 and after $2$ ns when the transients have settled. Here 
$J_{\rm IN} = 6~ {\rm MA/m}^2$ and $r$ is fixed at 0.01 m.
Note that in fig.~\ref{fig:EzImposed}, the field is due to
the applied potential alone while in fig.~\ref{fig:Ez} the
field is a superposition of the applied field and the 
electromagnetic field generated by the charges. 
$E_z$ changes little after acquiring the form 
shown in fig.~\ref{fig:Ez} characterized by its nearly vanishing 
value at the cathode. 

The radial profile of the current density for 
$J_{\rm IN} = 6~ {\rm MA/m}^2$ is shown in fig.~\ref{fig:Jz_radial}.
The plotted value of the z component of the transmitted current 
density, $J_z$, 
is averaged over 10 ns and the section is
taken close to the anode. The variation of the radially averaged
current density with time is shown in fig.~\ref{fig:Jz_t}. In 
order to reduce fluctuations, a local time averaging has also 
been performed.

\begin{center}
\begin{figure}[tbp]
\hspace*{0.1cm}\epsfig{figure=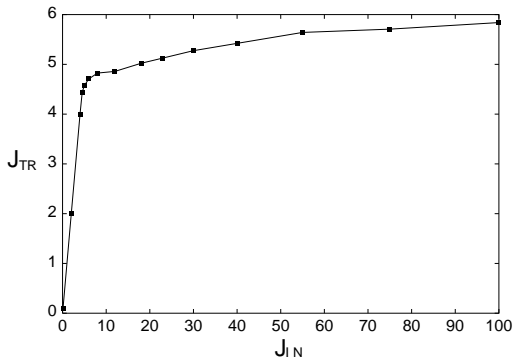,width=4.5cm,angle=270}

\vskip 0.25 in
\caption[ty] {Time averaged transmitted current density plotted
against the injected current density for the axially symmetric 
diode with $V = 250$ kV. Both are measured in ${\rm MA/m^2}$.  
Unlike the 1-dimensional case, 
no saturation can be observed for $J_{\rm IN} > J_{\rm CR}$.}

\label{fig:10}
\end{figure}
\end{center}

\begin{center}
\begin{figure}[tbp]
\hspace*{0.1cm}\epsfig{figure=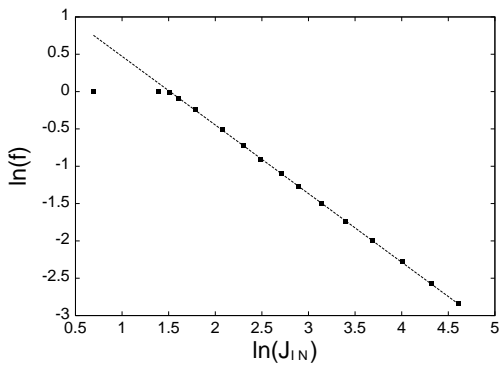,width=4.5cm,angle=270}
\vskip 0.25 in
\caption[ty] {A log-log plot of the average fraction of transmitted
electrons versus the injected current for the axially symmetric 
diode with $V = 250~{\rm kV}$. 
The linear fit for $J_{\rm IN} > 5$ confirms that
$f \sim J_{\rm IN}^{-\beta}$ with $\beta = 0.92$.}
\label{fig:11}
\end{figure}
\end{center}

We now present the central result of this communication.
For the non-relativistic studies, the potential considered was
$250$ kV for which $J_{\rm CL}(1)=4.57~{\rm MA/m^2}$ and
$J_{\rm CL}(2)=1.0354~J_{\rm CL}(1) = 4.73~{\rm MA/m^2}$. 
The injected
current was varied in the range $0 \le J_{\rm IN} < 100$. While 
criticality sets in at around $J_{\rm CL}(1)$, there appears to be
no saturation in this range (see Fig.~10). While $J_{\rm TR}
\simeq J_{\rm IN}$ for $J_{\rm IN} < J_{\rm CR}$, the transmitted
current increases as a power law beyond the critical current. This is
evident from a plot of the average fraction of electrons reaching the
anode as a function of the injected current (Fig.~11). The power law
behaviour is evident with $f \sim J_{\rm IN}^{-0.92}$. Thus the
transmitted current increases beyond the critical current as
$J_{\rm TR} = fJ_{\rm IN} \sim J_{\rm IN}^{0.08}$.

\begin{center}
\begin{figure}[tbp]
\hspace*{0.1cm}\epsfig{figure=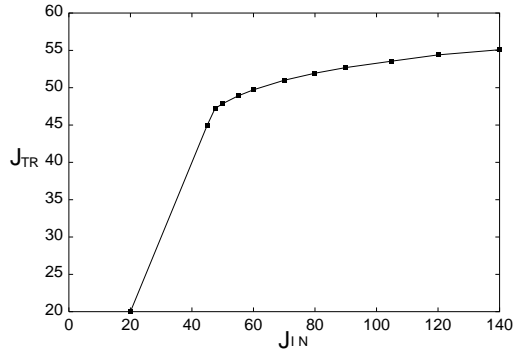,width=4.5cm,angle=270}
\vskip 0.25 in
\caption[ty] {Time averaged transmitted current density plotted
against the injected current density for $V = 2$ MV.
Both are measured in ${\rm MA/m^2}$.}
\label{fig:12}
\end{figure}
\end{center}

\begin{center}
\begin{figure}[tbp]
\hspace*{0.1cm}\epsfig{figure=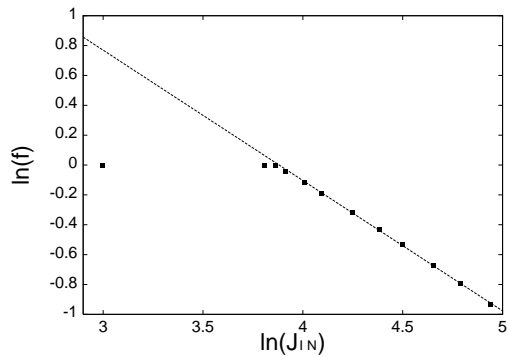,width=4.5cm,angle=270}
\vskip 0.25 in
\caption[ty] {A log-log plot of the average fraction of transmitted
electrons versus the injected current for V=2 MV. 
The fit is made for $J_{\rm IN} > 50$ and the value of $\beta$ is 0.873.}
\label{fig:13}
\end{figure}
\end{center}

	A similar study for a relativistic case (2 MV)
leads us to an identical conclusion - the transmitted current
increases beyond the critical current as evident from
Fig.~12. The power law behaviour can be seen from a plot of
${\rm ln}(f)$ vs ${\rm ln}(J_{\rm IN})$ (Fig.~13) and for this particular case,
$\beta=0.873$. Thus, $J_{\rm TR} \sim J_{\rm IN}^{0.127}$. Note that
in the relativistic case, there is a strong pinching effect due
to the self magnetic field. The observed value of the critical 
current, $J_{\rm CR}$, is  $47~{\rm MA/m^2}$
which is much less than the 1-dimensional prediction ($80~{\rm MA/m^2}$).

	These results do not preclude the existence of asymptotic
saturation. Rather, a fit of $J_{\rm TR}$ using the form
$J_{\rm SAT} - A/J_{\rm IN}^\gamma$ provides a marginally better
fit than the power law. However, a unique set of values for
($J_{\rm SAT},A, \gamma$) is hard to determine with the existing
data as several sets exist with nearly the same error (sum of squares).
Nevertheless, it is clear from these studies 
that $J_{\rm TR}$ does not saturate for finite values of injected currents 
beyond the critical value.

\section{Conclusions}
The existence of a limiting current in an
axially symmetric cavity diode has been investigated.
It is found that
$J_{\rm TR}$ does not saturate with an increase in $J_{\rm IN}$
beyond the critical value characterised by the onset of reflection.
This is unlike the case of one-dimensional diode which exhibits
saturation of $J_{\rm TR}$ as soon as $J_{\rm IN}$ exceeds
$J_{\rm CR}$. 
Our main result is the power law behaviour,
established through numerical computations 
between the transmitted current $J_{\rm TR}$ and finite values of the 
injected current $J_{\rm IN}$ in an axially symmetric 
diode operating above the critical current
$J_{\rm CR}$. The exponent depends on the voltage.
We have also explored the possibility of asymptotic 
saturation and found that there exist sets of parameters
($J_{\rm SAT},A, \gamma$) for which the form
$J_{\rm TR} = J_{\rm SAT} - A/J_{\rm IN}^\gamma$ provides 
a marginally better fit than the power law 
$J_{\rm TR} \sim J_{\rm IN}^\beta$.

The findings here thus have an important bearing 
on numerical simulations which generally assume saturation of
transmitted current even in two-dimensional diodes at a {\em finite}
value of the injected currenbt. Analytical and
numerical studies of a one-dimensional diode have also been
presented for the sake of comparison and completeness.

\newcommand{\PRL}[1]{{Phys.\ Rev.\ Lett.}\/ {\bf #1}}
\newcommand{\PR}[1]{{Phys.\ Rev.}\/ {\bf #1}}
\newcommand{\PRA}[1]{{Phys.\ Rev.\ A}\/ {\bf #1}}
\newcommand{\PRB}[1]{{Phys.\ Rev.\ B}\/ {\bf #1}}
\newcommand{\PRD}[1]{{Phys.\ Rev.\ D}\/ {\bf #1}}
\newcommand{\PRE}[1]{{Phys.\ Rev.\ E}\/ {\bf #1}}
\newcommand{\PP}[1]{{Phys.\ Plasmas\ }\/ {\bf #1}}
\newcommand{\JAP}[1]{{J.\ App.\ Phys. }\/ {\bf #1}}
\newcommand{\ProcA}[1]{{Proc.\ R.\ Soc.\ London Ser.\ A}\/ {\bf #1}}
\newcommand{\JPA}[1]{{J.\ Phys.\ A}\/ {\bf #1}}
\newcommand{\JPB}[1]{{J.\ Phys.\ B}\/ {\bf #1}}
\newcommand{\JCP}[1]{{J.\ Chem.\ Phys.}\/ {\bf #1}}
\newcommand{\JPC}[1]{{J.\ Phys.\ Chem.}\/ {\bf #1}}
\newcommand{\JMP}[1]{{J.\ Math.\ Phys.}\/ {\bf #1}}
\newcommand{\JSP}[1]{{J.\ Stat.\ Phys.}\/ {\bf #1}}
\newcommand{\AP}[1]{{Ann.\ Phys.}\/ {\bf #1}}
\newcommand{\PLB}[1]{{Phys.\ Lett.\ B}\/ {\bf #1}}
\newcommand{\PLA}[1]{{Phys.\ Lett.\ A}\/ {\bf #1}}
\newcommand{\PD}[1]{{Physica D}\/ {\bf #1}}
\newcommand{\NPB}[1]{{Nucl.\ Phys.\ B}\/ {\bf #1}}
\newcommand{\INCB}[1]{{Il Nuov.\ Cim.\ B}\/ {\bf #1}}
\newcommand{\JETP}[1]{{Sov.\ Phys.\ JETP}\/ {\bf #1}}
\newcommand{\JETPL}[1]{{JETP Lett.\ }\/ {\bf #1}}
\newcommand{\RMS}[1]{{Russ.\ Math.\ Surv.}\/ {\bf #1}}
\newcommand{\USSR}[1]{{Math.\ USSR.\ Sb.}\/ {\bf #1}}
\newcommand{\PST}[1]{{Phys.\ Scripta T}\/ {\bf #1}}
\newcommand{\CM}[1]{{Cont.\ Math.}\/ {\bf #1}}
\newcommand{\JMPA}[1]{{J.\ Math.\ Pure Appl.}\/ {\bf #1}}
\newcommand{\CMP}[1]{{Comm.\ Math.\ Phys.}\/ {\bf #1}}
\newcommand{\PRS}[1]{{Proc.\ R.\ Soc. Lond.\ A}\/ {\bf #1}}


\begin{references}
\bibitem{drift} A similar analysis for drift space can be found in \cite{bb}.
\bibitem{transient} In each case, time averaging is performed after
eliminating transients.
\end{references}
\end{document}